\documentclass[pre,aps,twocolumn,showpacs]{revtex4}
\usepackage{amsmath}
\usepackage{graphics}
\usepackage{epsfig}
\begin{document}
\title{Directed $d$-mer diffusion describing Kardar-Parisi-Zhang type 
of surface growth}
\author{G\'eza \'Odor (1), Bartosz Liedke (2) and Karl-Heinz Heinig (2)}
\affiliation{(1)Research Institute for Technical Physics and Materials Science, 
P.O.Box 49, H-1525 Budapest, Hungary\\
(2) Institute of Ion Beam Physics and Materials Research,
Forschungszentrum Dresden - Rossendorf, 
P.O.Box 51 01 19, 01314 Dresden, Germany}
\begin{abstract}
We show that $d+1$-dimensional surface growth models can be mapped
onto driven lattice gases of $d$-mers. The continuous surface
growth corresponds to one dimensional drift of $d$-mers 
perpendicular to the $\left( d-1\right)$-dimensional "plane" 
spanned by the $d$-mers. This facilitates efficient, bit-coded 
algorithms with generalized Kawasaki dynamics of spins. 
Our simulations in $d=2,3,4,5$ dimensions provide scaling exponent 
estimates on much larger system sizes and simulations times published so far,
where the effective growth exponent exhibits an increase.
We provide evidence for the agreement with field theoretical predictions 
of the Kardar-Parisi-Zhang universality class and numerical results. 
We show that the $\left(2+1\right)$-dimensional exponents conciliate 
with the values suggested by L\"{a}ssig within error margin,
for the largest system sizes studied here, but we can't support his 
predictions for $\left(3+1\right)d$ numerically.
\end{abstract}
\maketitle

One of the simplest nonlinear stochastic differential equation
set up by Kardar, Parisi and Zhang (KPZ) \cite{KPZeq} describes the 
dynamics of growth processes in the thermodynamic limit. 
It specifies the evolution of the height function $h(\mathbf{x},t)$ in the
$d$ dimensional space
\begin{equation}  \label{KPZ-e}
\partial_t h(\mathbf{x},t) = v + \sigma\nabla^2 h(\mathbf{x},t) +
\lambda(\nabla h(\mathbf{x},t))^2 + \eta(\mathbf{x},t) \ .
\end{equation}
Here $v$ and $\lambda$ are the amplitudes of the mean and local growth
velocity, $\sigma$ is a smoothing surface tension coefficient and $\eta$
roughens the surface by a zero-average Gaussian noise field exhibiting 
the variance 
$\langle\eta(\mathbf{x},t)\eta(\mathbf{x^{\prime}},t^{\prime})\rangle = 2 D
\delta^d (\mathbf{x-x^{\prime}})(t-t^{\prime})$. 
The notation $D$ is used for the noise amplitude and $\langle\rangle$ 
means the distribution average.

The KPZ equation was inspired in part by the the stochastic Burgers
equation \cite{Burgers74}, which belongs to the same universality class 
\cite{forster77}, and it became the subject of many theoretical studies
\cite{HZ95,barabasi,krug-rev}. Besides, it models other important physical 
phenomena such as directed polymers \cite{kardar85}, randomly stirred fluid 
\cite{forster77}, dissipative transport \cite{beijeren85,janssen86},
and the magnetic flux lines in superconductors \cite{hwa92}. 
The equation is solvable in $\left( 1+1\right) d$ \cite{kardar87},
but in higher dimensions approximations are available only. 
As the result of the competition of roughening and smoothing terms, 
models described by the KPZ equation exhibit a roughening phase transition
between a weak-coupling regime ($\lambda <\lambda _{c}$), governed by the
Edwards-Wilkinson (EW) fixed point at $\lambda =0$ \cite{EWc}, and a strong
coupling phase. The strong coupling fixed point is inaccessible by
perturbative renormalization group (RG) method. 
Therefore, the KPZ phase space has been the subject of controversies 
and the value of the upper critical dimension has been debated for 
a long time.

Using a directed polymer representation, the validity of a scaling 
hypothesis \cite{DK92} and the two-loop RG calculation for $d\ge 2$
\cite{FT94} was confirmed and extended to all orders in $d=(2+\epsilon)$ 
\cite{L95}. These results provided an argument for an upper 
critical dimension $d_{c}=4$ of the roughening transition, but
the strong-coupling rough phase is not accessible by perturbation
theory. Above $d=1$ the scaling behavior in the rough phase has been very 
controversial, diverse values for the scaling exponents were claimed 
\cite{H90,Ste94,L98}. In particular, assuming that height correlations 
exhibit no multiscaling and satisfy an operator product expansion, exact 
field-theoretic methods lead to rational number growth values
in two and three dimensions \cite{L98}.
Some theoretical approaches predict that $d_{c}=4$ is an upper critical 
dimension of the rough phase \cite{M95,L97,B98,F05}.
Recently, a non-perturbative RG study has been able to describe the strong 
coupling fixed point and has provided indications for a possible qualitative 
change of the critical behavior around $d=4$ \cite{CDDW09}. This is in 
contradiction with the numerical results \cite{AHK93,ala99,MPP,MPPR02}, 
which predict the lack of an upper critical dimension. 

\begin{figure}
\begin{center}
\epsfxsize 70mm
\epsffile{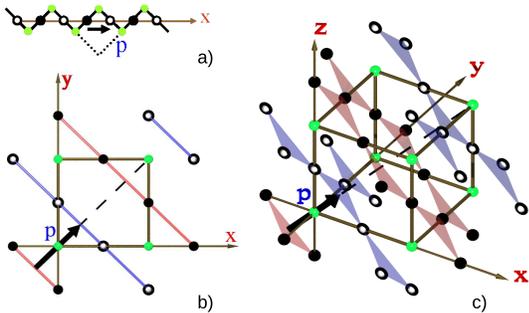}
\end{center}
\caption{(Color online) Mapping of one (a), two (b), three (c)
dimensional surface growth (discrete up/down derivatives) onto
$1,2,3$-dimensional oriented motion of $d$-mers. In general a
deposition event corresponds to a $d$-Kawasaki exchange, or a 
$d$-mer hopping in the bisectrix direction of axes. Up/down slopes
between lattice points (green dots) are denoted by full/empty 
circles, corresponding to particles/holes on the projection space.
The surface continuity translates into particle exclusion, 
which makes the model nontrivial in higher dimensions.}
\label{map}
\end{figure}

Mapping of surface growth onto reaction-diffusion system 
allow effective numerical simulations \cite{dimerlcikk,Orev}.
As a generalization of the $1+1$ dimensional roof-top model
\cite{kpz-asepmap,meakin} and the $2+1$ dimensional octahedron model 
\cite{asep2dcikk}, we consider the deposition and removal processes 
of higher dimensional objects on $d\ge 2$ dimensional surfaces. 
We remind that in $1+1$ dimensions a continuous surface line having
no overhangs can be approximated by 45 degree up/down slope elements 
(after appropriate length rescaling). A process with KPZ scaling 
can be realized by deposition at local minima or removal of local 
maxima (roof-top).
If we associate the up slopes with 'particles' and down slopes with 
'holes' of the base lattice (see Fig.~\ref{map}.a), the 
adsorption/desorption corresponds to the asymmetric exclusion process (ASEP) 
\cite{Ligget}. This is a lattice gas \cite{KLS84,S-Z}, where particles can 
hop on adjacent sites with asymmetric rates and hard-core exclusion.
Its behavior is well known, and variations of ASEP (disorder,
interactions ... etc.) correspond to variations of $1+1$ dimensional
KPZ growth models.

We have extended the roof-top construction to $\left(2+1\right)$ 
dimensions \cite{asep2dcikk} by the introduction of octahedra
having four slopes.
The up edges in the $x$ or $y$ directions can be 
represented by '$+1$', while the down ones by '$-1$', and a surface 
element update is a generalized (Kawasaki) exchange 
(Eq.~(3) of \cite{asep2dcikk}). The translation of up edges to 
'particles' and the down ones to 'holes' of the base lattice 
maps particle deposition/removal processes onto two simultaneous 
particle moves (one in the $x$ and one in the $y$ direction). 
One can also consider it as a dimer move in the bisectrix direction 
of $x$ and $y$  (see Fig.~\ref{map}.b). 
Therefore, the  $\left(2+1\right)$ surface dynamics can be mapped 
onto a "two-dimensional ASEP" of oriented dimers exhibiting hard-core 
exclusion. The asymmetric drift corresponds to an evolving 
surface exhibiting KPZ scaling, while the symmetric dimer diffusion 
is related to the EW behavior.  

Now we proceed with this kind of construction, considering
the discrete slope variables in higher dimensions and generalize the
simultaneous $+1 \leftrightarrow -1$ (Kawasaki) exchange rule of 
them (Eq.(3) of \cite{asep2dcikk}) to $d$-dimensional updates
\begin{equation}
\left( 
\begin{array}{cc}
-1 & 1 \\ 
-1 & 1 \\ 
-1 & 1 \\ 
... & 
\end{array}
\right) \overset{p}{\underset{q}{\rightleftharpoons }}\left( 
\begin{array}{cc}
1 & -1 \\ 
1 & -1 \\ 
1 & -1 \\ 
... & 
\end{array} 
\right)  \ , \label{rule}
\end{equation}
with probability $p$ for attachment and probability $q$ for detachment
(see Fig.~\ref{map}.c for the 3d case).
It is well known \cite{barabasi} that the surface evolution of the 
deterministic KPZ growth are described also by the Burgers equation 
\cite{Burgers74} for growth velocities \textbf{v}(\textbf{x},t) 
in the surface normal obeying
\begin{equation}  \label{BUR-e}
\partial_t \mathbf{v}(\mathbf{x},t) = \sigma\nabla^2 \mathbf{v}(\mathbf{x}%
,t) + \lambda \mathbf{v}(\mathbf{x},t) \nabla \mathbf{v}(\mathbf{x},t) 
\end{equation}
due to the transformation $\mathbf{v}(\mathbf{x},t) = \nabla h(\mathbf{x},t)$.

In the forthcoming part we will prove that our microscopic model for $d$-mers 
in the continuum limit can be mapped onto the anisotropic version of 
Eq.~(\ref{BUR-e}), similarly as shown in lower dimensions
\cite{kpz-asepmap,asep2dcikk}. The derivation is based on the 
formulation of the reduction of possible updates.
Our surface model is represented by the discrete derivative elements: 
$\delta _{x}$, $\delta _{y}$, $\delta _{z}$ ... ($\in \pm 1$) at every 
lattice points. A generalized Kawasaki update (\ref{rule}) is defined 
by a matrix
\begin{eqnarray} \label{up1}
\left( 
\begin{array}{cc}
\delta _{x}(i-1,j,k,...) & \delta _{x}(i,j,k,...) \\ 
\delta _{y}(i,j-1,k,...) & \delta _{y}(i,j,k,...) \\ 
\delta _{z}(i,j,k-1,...) & \delta _{z}(i,j,k,...) \\ 
... & 
\end{array}
\right) \ .
\end{eqnarray}
In $d$ dimensions we define vectors of the slopes,
the columns of (\ref{up1}), analogously to one and two dimensions: 
$\overline{\sigma }_{i,j,k,..}=(\delta _{x}(i-1,j,k,...),\delta
_{y}(i,j-1,k,...),...)\ ,  \label{svect}$ around the lattice
point, which we select for deposition/removal update
and set up a microscopic master equation
\begin{eqnarray}
\partial _{t}P(\{\overline{\sigma }\},t)
&=&\sum_{i,j,k,...}w_{i,j,k,...}^{\prime }(\{\overline{\sigma }\})P(\{%
\overline{\sigma }^{\prime }\},t)  \notag  \label{mastereq} \\
&-&\sum_{i,j,k,...}w_{i,j,k,...}(\{\overline{\sigma }\})
P(\{\overline{\sigma }\},t)
\end{eqnarray}%
with the probability distribution $P(\{\overline{\sigma }\},t)$. Here the
prime index denotes the state of $\overline{\sigma}$ following the update
(\ref{rule}). The transition probability of $\overline{\sigma}$-s
can be expressed as
\begin{eqnarray} \label{wtr}
w_{i,j,k,...}(\{\overline{\sigma }\}) &=&A[1-\overline{\sigma }%
_{i+1,j+1,k+1,...}\overline{\sigma }_{i,j,k,...}  \label{mast} \\
&+&\lambda (\overline{\sigma }_{i+1,j+1,k+1,...}-\overline{\sigma }%
_{i,j,k,...})]\ ,  \notag  
\end{eqnarray}
with $\lambda = 2 p/(p+q)-1$ parametrization, which formally looks like
the Kawasaki exchange probability in $1$d, except the factor $A$,
which is necessary to avoid surface discontinuity creation in higher
dimensions. This means that we update the slope configurations only if the
values of all coordinates of the vector $\overline{\sigma}$ are identical
as shown by (\ref{rule}). One can allow formally these updates 
via the expression
\begin{eqnarray}\label{filt}
A &=&1/2^{d+1}\det [(\overline{\sigma }_{i,j,k,..}+C\overline{\sigma }%
_{i,j,k,..}) \\
&&\times (\overline{\sigma }_{i+1,j+1,k+1,..}+C\overline{\sigma }%
_{i+1,j+1,k+1,..})I]\ ,  \notag
\end{eqnarray}%
where $I$ and $C$ are the unity and the cyclic permutation matrices
respectively. The matrix $C$ shifts each coordinate value to the next 
index value. Thus for $\overline{\sigma}$-s with mixed coordinate values, 
the vectors 
$(\overline{\sigma}_{i,j,k,..} + C\overline{\sigma}_{i,j,k,..} )=\overline{k}$ 
or $(\overline{\sigma}_{i+1,j+1,k+1,..} +C\overline{\sigma}_{i+1,j+1,k+1,..} )
=\overline{k'}$ possess zero elements.
Therefore the determinant of $\overline{k} \overline{k'} I$,
being the product of the diagonal elements, is zero in case of 
mixed coordinates and $A=1/2^{d+1}$ in case of equal coordinates. 

For example a $d=3$ update is prohibited when the slope vector is 
$\overline{\sigma }=(1,1,-1)$, because $\overline{k}$ has one coordinate 
value of zero
\begin{equation*}
\overline{k}=\left( 
\begin{array}{c}
1 \\ 
1 \\ 
-1%
\end{array}%
\right) + \left( 
\begin{array}{ccc}
0 & 1 & 0 \\ 
0 & 0 & 1 \\ 
1 & 0 & 0%
\end{array}%
\right) \left( 
\begin{array}{c}
1 \\ 
1 \\ 
-1%
\end{array}%
\right) =\left( 
\begin{array}{c}
2 \\ 
0 \\ 
-2%
\end{array}
\right) \ .
\end{equation*}

By calling '$+1$'-s as particles and the '$-1$'-s as holes of the base
lattice, their synchronous update can be considered to be a single step 
motion of an oriented $d$-mer in the bisectrix direction of the 
$x$, $y$, $z$, ... coordinate axes. Thus $d$-mers follow one-dimensional 
kinetics, described by Kawasaki exchanges (\ref{mast}).
To obtain a one-to-one mapping we update neighborhoods of the 
lattice points denoted by the green dots of Fig.~\ref{map}.
 
To derive Eq.~(\ref{BUR-e}) first we have to average over the slope vectors 
\begin{equation}
\langle \overline{\sigma }\rangle =\sum_{\{\overline{\sigma }\}} 
\overline{\sigma } P(\{\overline{\sigma }\},t) .
\end{equation}
By calculating its time derivative using the master equation
(\ref{mastereq}) and the transition probabilities (\ref{wtr})
\begin{eqnarray}
&&\partial _{t}\langle \overline{\sigma} \rangle =
\sum_{\{\overline{\sigma }\}} \Big[ \overline{\sigma} 
\sum_{i,j,k,...}w_{i,j,k,...}^{\prime }(\{\overline{\sigma }\})
P(\{\overline{\sigma }^{\prime }\},t) \notag \\
&-&\overline{\sigma} \sum_{i,j,k,...}w_{i,j,k,...}
(\{\overline{\sigma }\})P(\{\overline{\sigma }\},t) \Big] \ ,
\end{eqnarray} 
in which we filter out vectors of non-equal coordinates (\ref{filt})
(thus $w_{i,j,k,...}$ is nonzero only if
$\overline{\sigma}_{i,j,...}\ne\overline{\sigma}_{i+1,j+1,...}$)
we can obtain 
\begin{eqnarray}
&&2\partial _{t}\langle \overline{\sigma }_{i,j,k,...}\rangle =\langle 
\overline{\sigma }_{i-1,j-1,...}\rangle -2\langle \overline{\sigma }%
_{i,j,...}\rangle  \\
&+&\langle \overline{\sigma }_{i+1,j+1,...}\rangle +\lambda \langle 
\overline{\sigma }_{i,j}(\overline{\sigma }_{i+1,j+1,...}-\overline{%
\sigma }_{i-1,j-1,...})\rangle \ ,  \notag
\end{eqnarray}
analogously to one dimension \cite{kpz-asepmap}.
Here one can see the discrete first and second differentials of 
$\overline{\sigma }_{i,j,k,...}$ corresponding to the operators 
of Eq.~(\ref{BUR-e}) in the bisectrix direction of the axes. These 
differentials are one-dimensional, because the $d$-mer dynamics 
is one-dimensional. 
In principle one could derive a set of coupled Burgers equations 
for the particles in each direction in an isotropic way in 
accordance with isotropic surface model, but the coordinated 
movements reduce the dimensionality and we can map onto an 
anisotropic equation of $d$-mers.

Making a continuum limit in each direction and taking into account 
the relation of height and slope variables
($\mathbf{v}(\mathbf{x},t) = \nabla h(\mathbf{x},t)$),
we can arrive to the deterministic KPZ equation.
The nonlinear term vanishes for $p=q$ ($\lambda =0$). 
The sign of the coefficient $\lambda $ of the nonlinear term can be 
interpreted as follows: For $p>q$ positive non-linearity 
(positive excess velocity) it is a consequence of growth with voids.

Since this derivation was applied just for the first one
in the hierarchy of equations for correlation functions
it does not prove the equivalence to the stochastic KPZ.
Furthermore, the form of the noise term, which was not
considered in our derivation, may also introduce differences. 
Although our surface model is spatially isotropic, we can map it
onto a one-dimensional Burgers equation (of extended objects), 
therefore anisotropic scaling behavior might be expected.
However, by going into the continuum description the hard-core
exclusions necessary to provide continuous surfaces is lost, and
the resulting equation looks trivial.
 
Here we investigate by numerical simulations this isotopic surface
growth model via the one-dimensional directed migration of $d$-mers 
in the $d$-dimensional space. We have developed bit-coded algorithms 
for the updates (\ref{rule}) and run it with $p=1$, such that randomness 
comes from the site selection only. Therefore it is important to use 
a very good random number generator, which provides uniformly 
distributed numbers with high resolution. Otherwise we would realize 
a process with quenched disorder, which for KPZ belongs to a 
different universality class (see \cite{Orev}). 
We used the latest Mersenne-Twister generator \cite{MT} in general, 
which has very good statistical properties and which is very fast, 
especially in the SSE2 registers. We tested our results using other 
random number generators as well.
In practice each update site can be characterized by the $2^{d^2}$ 
different local slope configurations. However, due to the surface continuity
we need only a few bits of a world (1 byte for $d=2,3,4$
and two bytes for $d=5$) for this purpose. This allows an
efficient storage management in the computer memory 
and permits simulations of larger system sizes.  
The updates can be performed by logical 
operations, either on multiple samples at once, or on multiple 
(not overlapping) sites at once. Our bit-coded algorithm proved 
to be $\sim 40$ times faster than the conventional FORTRAN 90 code
we started with.
It is important to note that this stochastic cellular automaton
like representation of the surface growth opens the possibility
for an implementation on extremely fast graphic cards with massively
parallel processors. Furthermore the construction permits the extension
of the mapping for more complex surface reactions \cite{patscalcikk}.

We performed dynamical simulations by starting from stripe ordered 
particle distributions. This corresponds to a flat surface with a 
small intrinsic width. The considered lattices gases had the maximum 
linear sizes $L_{max}=2^{15}, 2^{10}, 2^8, 2^6$ for $d=2,3,4,5$ dimensions, 
respectively and periodic boundary conditions were applied. 
A single step of the lattice gas algorithm comprises a random site selection 
and in case of an appropriate neighborhood configuration a $p=1$
Kawasaki $d$-mer update (\ref{rule}).
The time is incremented by $1/L^d$ in units of Monte Carlo steps (MCs).
Throughout the paper we will use this unit of time.

We could exceed by magnitudes of order all previous numerical system 
sizes and simulation times. 
For example the largest five-dimensional simulations were done 
for $L=30$ and $t_{m}=230$ MCs \cite{AHK93}. 
Our $L=64$ simulations, where we have the good bulk/surface 
ratio: $5.4$, required 2GB memory size and a couple of weeks 
for a single realization up to $t_{max}=5000$ MCs.
Similarly, the largest sized simulations in $d=2$ for $L=11520$ 
system could achieve $t_{max}=10^4$ MCs \cite{TFW92}.
Our largest $L=32768$ sized simulations reached $t_{max}=44600$ MCs.
The longest runs for $L=4096$ passed the saturation at 
$t\simeq 4\times10^5$ MCs and the samples were followed up to 
$t_{max}=10^6$ MCs.

We run the these lattice gas simulations for $10-1000$ independent 
realizations for each dimension and size considered, and calculated 
$h_{x,y,..}(t)$ and the second moment
\begin{equation}\label{width}
W(L,t) = \Bigl[ \frac{1}{L^{2d}} \, \sum_{x,y,..}^L 
\,h^2_{x,y,...}(t) - \Bigl(\frac{1}{L^d} \, \sum_{x,y,...}^L \,h_{x,y,...}(t) 
\Bigr)^2 \Bigr]^{1/2} 
\end{equation}
from the height differences at certain sampling times. The growth is expected 
to follow the Family-Vicsek scaling~\cite{family} asymptotically, but due 
to the corrections it can be described by a power series
\begin{equation}\label{betacorr}
W(t,L\to\infty) = b t^{\beta}(1 + b_0 t^{\phi_0} + b_1 t^{\phi_1} ...) \ ,
\end{equation}
with the surface growth exponent $\beta$. For finite system, when the
correlation length exceeds $L$, the growth crosses over to a saturation
with the scaling law
\begin{equation}\label{alphacorr}
W(t\to\infty,L) = a L^{\alpha}(1 + a_0 L^{\omega_0} + a_1 L^{\omega_1} ...) \ ,
\end{equation}
characterized by the roughness exponent $\alpha$.
In our case the intrinsic width of the initial state, which is represented 
by a zig-zag surface of width $1/2$ (see Fig.\ref{map}), results in 
a constant correction term. Thus we have $b_0=1/2$, $\phi_0=-\beta$
and $a_0=1/2$, $\omega_0=-\alpha$.
During our scaling analysis we dropped this contribution by subtracting 
$W^2(0)=1/4$ from the raw data and consider the next leading order 
correction as leading one. Furthermore we disregarded the initial
time region $t < t_0 \simeq 50$, when basically an uncorrelated random 
deposition occurs.
The dynamical exponent $z$ can be expressed by the ratio $z = \alpha/\beta$ 
and in case of the Galilean invariance of an isotropic KPZ equation 
the $z=2-\alpha$ relation should also hold.

Besides the extensive simulations we have performed careful correction to 
scaling analysis by calculating the local slopes of the exponents. 
The effective exponent of the surface growth can be estimated similarly 
as in case of other scaling laws \cite{Orev}, as the discretized, 
logarithmic derivative of (\ref{width})
\begin{equation}  \label{beff}
\alpha_{eff}(L) = \frac {\ln W(t\to\infty,L) - \ln W(t\to\infty,L')} 
{\ln(L) - \ln(L^{\prime})} \ .
\end{equation}
It was determined numerically for different discretizations: $t/t^{\prime}=2,3$, and we tried to fit it with the leading-order correction ansatz, which can 
easily be deduced from (\ref{betacorr}) (see \cite{AA04} or \cite{Orev})
\begin{equation}  \label{betaslfit}
\beta_{eff}(t) = \beta + b_1\phi_1 t^{\phi_1} \ ,
\end{equation}
for $t>t_0$ and before the saturation region.
In other cases, such as ballistic deposition, which has a large unknown 
intrinsic width one can use another effective roughness exponent
definition introduced in \cite{AA04}. 

\begin{figure}
\begin{center}
\epsfxsize 70mm \epsffile{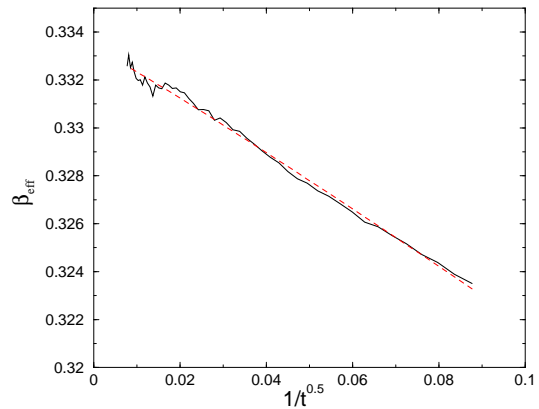}
\end{center}
\caption{(Color online) Effective exponents of the growth of the $d=1$
dimensional model. The solid line corresponds to the simulation result.
The dashed lines shows a fitting with the form (\ref{betaslfit}).}
\label{beta1}
\end{figure}

We tested our method with the {\it one-dimensional}, exactly known case.
Simulations were run on $L=5\times 10^5$ sized system up to $t_{max}=16666$ 
MCs for $40$ independent realizations. We determined the effective exponents 
$\beta_{eff}(t)$, which approaches $\beta=1/3$ from below, in a perfect 
agreement with the leading-order correction form (see Figure~\ref{beta1}). 
The fitting with (\ref{betaslfit}) on the local slopes data resulted in 
$\beta=0.333(5)$ and $\phi_1=-0.53$.

\begin{figure}
\begin{center}
\epsfxsize 70mm \epsffile{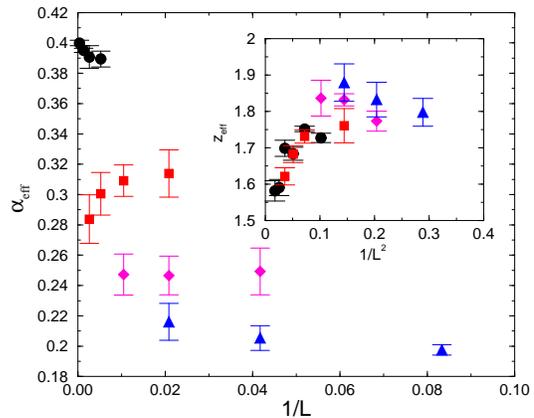}
\end{center}
\caption{(Color online) Local slopes of the finite size scaling of 
the saturation width in $d=2,3,4,5$ dimensions (top to bottom). 
Inset: Effective exponents of the characteristic times. }
\label{alphas}
\end{figure}

Similarly to the time dependence we can analyze the size dependence
following the saturation by determining the effective exponent of 
the roughness, which can be defined as the logarithmic derivative of 
(\ref{width})
\begin{equation}  \label{aeff}
\alpha_{eff}(L) = \frac {\ln W(t\to\infty,L) - \ln W(t\to\infty,L')}
{\ln(L) - \ln(L^{\prime})} \ .
\end{equation}
The finite size scaling was done for systems of linear sizes in between 
$L_{max}$ (discussed earlier) and $L_{min}$, which was $2^6$ for 2d, 
$2^5$ for 3d, $2^4$ for 4d and $2^3$ for 5d, respectively.
To handle the boundary conditions effectively, system sizes of power of 2
were simulated.
To get the asymptotic values we took into account all effective
exponent points shown on Fig.~\ref{alphas} and applied a leading order, 
linear fitting.
The error margins of exponents are estimated from the error-bars of 
Fig.~\ref{alphas}. This method gives a better estimate for the asymptotic 
values than just a least-square fitting on the data points, which 
completely disregards corrections to scaling.
We also calculated rough, but independent estimates for $z$ 
by measuring the relaxation time, i.e. the time needed to reach 
90\% of the saturation value. The asymptotic value is extrapolated
by a linear fitting: $z_{eff}(t) = z + c_1/L$.

\begin{figure}
\begin{center}
\epsfxsize 70mm \epsffile{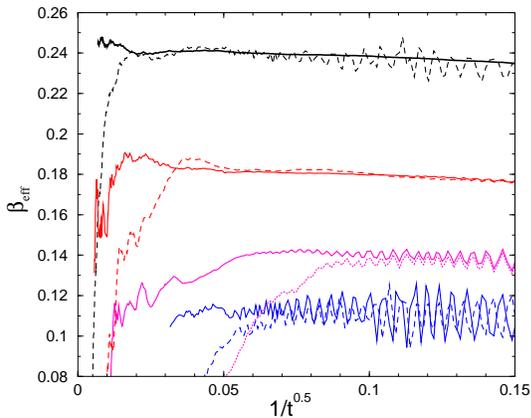}
\end{center}
\caption{(Color online) Effective exponents of the growth of the $d=2,3,4,5$
dimensional model (top to bottom). Solid lines correspond to the largest 
system $L=2^{15}$, $L=2^{10}$, $L=2^8$ and $L=2^6$ respectively. 
Dashed lines show our results of smaller sizes: $L=2^{10}$, $L=2^8$, $L=2^5$ 
and $L=2^5$ respectively, where saturation sets in earlier 
causing a cutoff in the scaling.}
\label{betas}
\end{figure}

In {\it two dimensions} we estimated the growth exponent in the
largest system sizes considered ($L=2^{15}, 2^{14}, 2^{13}$) 
(see Fig.~\ref{betas}). Fitting in the $50 < t < 44600$ time 
window with the form (\ref{betaslfit}) resulted in $b_1=0.83$
and $\beta=0.245(5)$, which is somewhat bigger than what was 
obtained by the largest known ($L=11520$) sized simulations:
$\beta=0.240(1)$ \cite{TFW92}, and all other previous numerical 
estimates including ours \cite{Ghai,AA04,asep2dcikk}.
This value conciliates with the $\beta=1/4$ RG exponent of \cite{L98}. 
One can obtain this value by the late time behavior of effective exponent, 
which has not been seen before, because finite size effects have screened it. 
On the graph one can see strong oscillations for 
$L=2^{10}$ and intermediate times, which are damped before saturation. 
In the one-dimensional ASEP model such oscillations are shown to be the 
consequence of density fluctuations being transported through a finite 
system by kinematic waves \cite{GMGB07}.
One can speculate that the slight final increase of $\beta_{eff}$ 
for the largest system sizes is just a fluctuation or oscillation effect, 
but we could not eliminate this overall tendency by increasing the statistics. 
Although the statistical fluctuations grow dramatically, as $t\to\infty$ 
the increase of the mean value is observable for each size $L > 2^{11}$.
Our error-bar of $\beta$ reflects this uncertainty.
The width saturation values have been investigated for $L=2^6, 2^7,...,2^{12}$. 
We took into account the leading order correction to-scaling by the following
Ansatz
\begin{equation}  \label{WFform}
\alpha_{eff}(L) = \alpha + a_1\omega_1 L^{\omega_1} \ ,
\end{equation}
but due to the larger error-bars we restricted it to a linear 
approximation: $\omega_1=-1$.
The local slopes of the steady state values $\alpha_{eff}(1/L)$ and of
$z_{eff}(1/L)$ are shown on Fig.~\ref{alphas}. 
This provides $\alpha=0.395(5)$ and $a_1=2.02$ for the roughness 
and $z_{eff}=1.58(10)$, with the linear coefficient $c_1=1.83$ for the 
dynamical exponent. This roughness exponent is in agreement with 
RG value \cite{L98}, and somewhat bigger than the existing figures 
$\alpha=0.393(3)$ \cite{MPP} for $L\le 1024$ and $\alpha=0.385(5)$
\cite{TFW92} for $L\le 128$.

In {\it three dimensions} the local slope analysis for $L=2^{10}$
results in $b_1 = 0.1$ and $\beta=0.184(5)$ agreeing with the numerical
results from the literature: $\beta=0.180(2)$ \cite{TFW92,AHK93}, 
$\beta=0.186(1)$ \cite{MPP}. But our estimate is much higher than 
$\beta=0.168(3)$ \cite{Ghai} (based on $L < 200$ sized simulations) 
and $\beta=1/6$ predicted by RG \cite{L98}. 
For the saturation we obtained $a_1=1.40$ and $\alpha=0.29(1)$, 
matching $\alpha=0.29$ of \cite{Ghai} and in marginal agreement 
with $\alpha=0.3135(15)$ of \cite{MPP} and $\alpha=0.308(2)$ of 
\cite{AHK93}. The direct $z_{eff}$ measurement exhibits a strong
correction to scaling: $z=1.60(1)$ ($c_1=1.10$) and one cannot 
differentiate it from the $2+1$ dimensional results within the 
error margins.

In {\it four spatial dimensions} our best fit for the growth exponent is
$b_1=1.08$ and $\beta=0.15(1)$. In the literature $\beta=0.16(1)$ 
\cite{ala99} and $\beta=0.146(1)$ \cite{MPP} values are reported.
For the width saturation values the linear fitting results 
in $\alpha=0.245(5)$ with $a_1=0.07$. 
This compares with the literature values $\alpha=0.245(1)$ \cite{AHK93} and 
$\alpha=0.255(5)$ \cite{MPP}.
The $z_{eff}$ seems to converge to $z=1.91(10)$ (with $c_1=-0.64$) 
but the fluctuations are very strong and we could not reach 
saturation for sizes larger than $L=128$. 
Going further by a factor of two in system sizes would require 
simulations with 8GB memory and very long CPU times.
Our results do not support the field theoretical prediction of $d_c=4$, 
because we don't observe the disappearance of power-law growth.

In {\it five dimensions} the local slopes suggest $b_1=0.134$
and $\beta=0.115(5)$ in agreement with $\beta=0.11(1)$ \cite{AHK93} 
reported for smaller sizes. One can find strong oscillations before the 
saturation regime. Again these are due to kinematic transport 
waves in finite system.
Initially for $L=64$ we saw a definite increase in $\beta_{eff}$ 
as $1/t < 0.005$ before the saturation, but this proved to be 
an artifact of the MT random number generator. When we used 
different, pseudo-random number generators: drand48 
\footnote{see for example: http://linux.die.net/man/3/drand48} 
or random() of language C, the growth tendency for very late times 
was much weaker. We think that the site selection, the only source of
randomness in case of $p=1$ might not be completely uniform 
among the $2^{30}$ possible places.
To confirm this we repeated the $5d$ simulations using $p=0.9$ with the
MT generator and found agreement with the results using drand48.
For the saturation we estimate $\alpha=0.22(1)$ with $a_1=0.08$ 
and $z=1.95(15)$ with $c_1=-0.55$.

In conclusion we have shown that the mapping of a KPZ surface growth model
onto driven lattice gases (DLG) can be extended to higher dimensions.
Although the growth of the surfaces exhibits the spatial symmetry of the
underlying lattice, one can map it onto an anisotropic DLG of more complex
objects. The coarse grained, continuum description of these $d$-mers 
is an anisotropic Burgers equation.
Still the DLG model is non-trivial, because it is just an oriented drift of 
$d$-mers with hard-core exclusions. The topological constraint is the 
consequence of the required surface continuity by the mapping. 
In two dimensions we confirmed \cite{patscalcikk} that the probability 
distribution $P(W^2)$ matches the universal scaling function determined 
for another KPZ model \cite{MPPR02}.
We presented effective bit-coding simulations and high precision results
for the exponents $\alpha$, $\beta$ and $z$ independently (see Table I.). 
The sensitive local slope analysis provides numerical agreement with 
former simulation results, but for larger sizes, which have not been 
investigated so far, we see a slight growing tendency in the 
$\beta_{eff}$ exponents in all dimensions. For $d=2$ our results 
marginally overlap with the $\beta=1/4$ value suggested some time ago by RG. 
The change towards a trivial behavior in higher dimensions in the 
DLG language would mean the disappearance of the topological 
constraints among the extended $d$-mer objects as they could 
follow a simple ASEP dynamics of point particles. This will be the 
target of further studies using massively parallel algorithms 
on graphic cards. We hope that we will be able to obtain a firm estimate
for the upper critical dimension using extrapolation techniques. 

\begin{table}[h]
\caption{Independent growth exponent estimates of the $d$-mer model
in different dimensions}
\begin{tabular}{|l|l|l|l|}
\hline
$d$   & $\alpha$ & $\beta$ & z \\
\hline
2     & 0.395(5) & 0.245(5) & 1.58(10) \\
3     & 0.29(1)  & 0.184(5) & 1.60(10) \\
4     & 0.245(5) & 0.15(1)  & 1.91(10) \\
5     & 0.22(1)  & 0.115(5) & 1.95(15) \\
\hline
\end{tabular}
\end{table}

We thank Zolt\'an R\'acz for the useful comments.
Support from the Hungarian research fund OTKA (Grant No. T77629), the
bilateral German-Hungarian exchange program DAAD-M\"OB (Grant Nos.
D/07/00302, 37-3/2008) is acknowledged. 
G. \'Odor thanks for the access to the Clustergrid and the NIIF supercomputer.

\bibliography{ws-book9x6}

\end{document}